\documentclass[epj]{svjour}
\usepackage{graphicx}
\input{epsf}
\usepackage{amssymb}
\begin{document}

\title{Google matrix of the world trade network} 

\author{L.Ermann \and D.L.Shepelyansky}
\institute{Laboratoire de Physique Th\'eorique du CNRS, IRSAMC, 
Universit\'e de Toulouse, UPS, 31062 Toulouse, France}

\date{March 25, 2011}


\abstract{Using the United Nations Commodity Trade Statistics 
Database \cite{comtrade} we construct the Google matrix
of the world trade network and analyze its properties
for various trade commodities for 
all countries and all available years from 1962 to 2009.
The trade flows on this network are classified with the
help of PageRank and CheiRank algorithms developed for
the World Wide Web and other large scale directed networks.
For the world trade this ranking treats all countries on 
equal democratic grounds independent of country richness. 
Still this method puts at the top a group of industrially 
developed countries for trade in {\it all commodities}.
Our study establishes the existence of two solid 
state like domains of rich and poor countries
which remain stable in time, while the majority
of countries are shown to be in a gas like phase with
strong rank fluctuations. A simple random matrix model
provides a good description of statistical distribution of countries
in two-dimensional rank plane. The comparison
with usual ranking by export and import
highlights new features and possibilities 
of our approach.  
}

\maketitle

\section{Introduction}
The analysis and understanding of world trade is of
primary importance for modern international economics \cite{krugman2011}.
Usually the world trade ranking of countries is done 
according their export and/or import counted in {\it USD} \cite{cia2010}.
In such an approach the rich countries naturally go
at the top of the listing simply due the fact that they
are rich and not necessary due to the fact that their trade network
is efficient, broad and competitive. 
In fact the trade between countries 
represents a directed network and hence it is  natural to apply 
modern methods of directed networks to analyze the
properties of this network. Indeed, on a scale of last decade
the modern society developed enormously large directed networks 
which started to play a very important role.
Among them we can list the World Wide Web (WWW),
Facebook, Wikipedia and many others.
The information retrieval and ranking of such large networks 
became a formidable challenge of modern society. 

An efficient approach to solution of this problem
was proposed in \cite{brin} on the basis
of construction of the Google matrix
of the network and ranking all its nodes 
with the help of the PageRank algorithm 
(see detailed description in \cite{googlebook}).
The elements $G_{ij}$ of the Google matrix of a network 
with $N$ nodes are defined as
\begin{equation}
  G_{ij} = \alpha S_{ij} + (1-\alpha) / N \;\; ,
\label{eq1} 
\end{equation} 
where the matrix $S$ is obtained by normalizing 
to unity all columns of the adjacency matrix $A_{i,j}$,
and replacing columns with only zero elements by $1/N$.
Usually for the WWW an element $A_{ij}$ of the adjacency matrix 
is equal to unity
if a node $j$ points to node $i$ and zero otherwise.
The damping parameter $\alpha$ in the WWW context describes the probability 
$(1-\alpha)$ to jump to any node for a random surfer. 
The value $\alpha = 0.85$  gives
a good classification for WWW \cite{googlebook}.
By construction the Google matrix belongs to the class
of Perron-Frobenius operators and Markov chains \cite{googlebook}, 
its largest eigenvalue 
is $\lambda = 1$ and other eigenvalues 
have $|\lambda| \le \alpha$.
According to the Perron-Frobenius theorem
the right eigenvector,  called the PageRank vector,
has maximal $\lambda=1$ and  non-negative
elements that have a meaning of probability
$P(i)$ attributed to node $i$.
Thus all nodes can be ordered in a decreasing
order of probability $P(i)$ with the corresponding
increasing PageRank index $K(i)$.
The presence of gap between $\lambda=1$ and 
$|\lambda|=\alpha$ ensures a convergence of 
a random initial vector to the PageRank
after about 50 multiplications by matrix $G$.
Such a ranking based on the PageRank algorithm
forms the basis of the Google search engine \cite{googlebook}.  
It is established that
a dependence of PageRank probability $P(i)$ on rank $K(i)$
is well described by a power law $P(K) \propto 1/K^{\beta_{in}}$ with
$\beta_{in} \approx 0.9$. This is consistent with the relation
$\beta_{in}=1/(\mu_{in}-1)$ corresponding to the average
proportionality of PageRank probability $P(i)$
to its in-degree distribution $w_{in}(k) \propto 1/k^{\mu_{in}}$
where $k(i)$ is a number of ingoing links for a node $i$  
\cite{litvak,googlebook}. For the WWW it is found that
for the ingoing links $\mu_{in} \approx 2.1$ (with $\beta_{in} \approx 0.9$)
while for out-degree distribution
$w_{out}$ of
outgoing links a power law has the exponent  $\mu_{out} \approx 2.7$
\cite{donato,upfal}. We note that  PageRank is used 
for ranking in various directed networks including
citation network of Physical Review 
\cite{redner,fortunato} and 
for rating of the total importance of scientific journals 
\cite{bergstrom}.
\begin{figure}[ht]
\begin{center}  
\includegraphics[width=.225\textwidth]{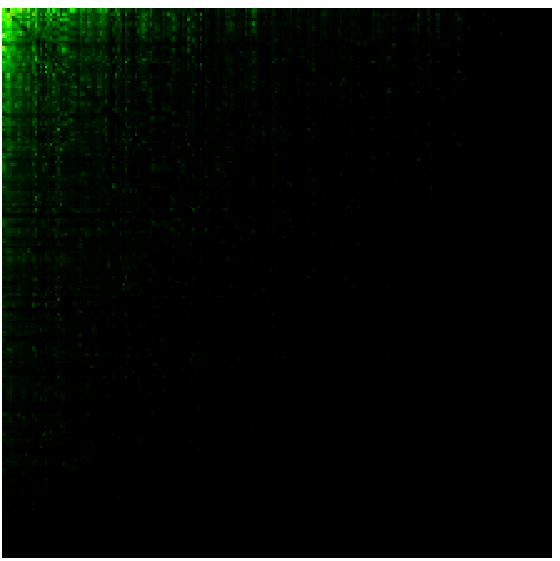} \hspace{0.01\textwidth}
\includegraphics[width=.225\textwidth]{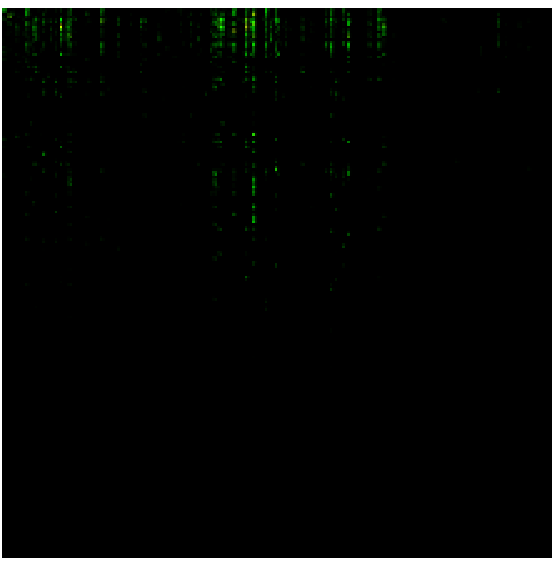} \\
\vspace{0.01\textwidth}
\includegraphics[width=.225\textwidth]{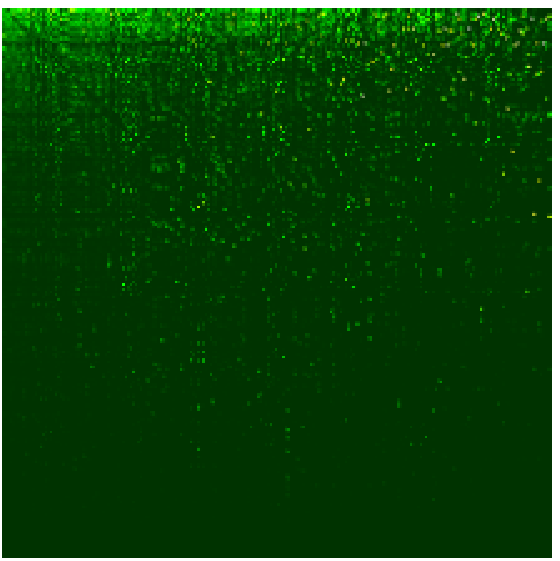} \hspace{0.01\textwidth}
\includegraphics[width=.225\textwidth]{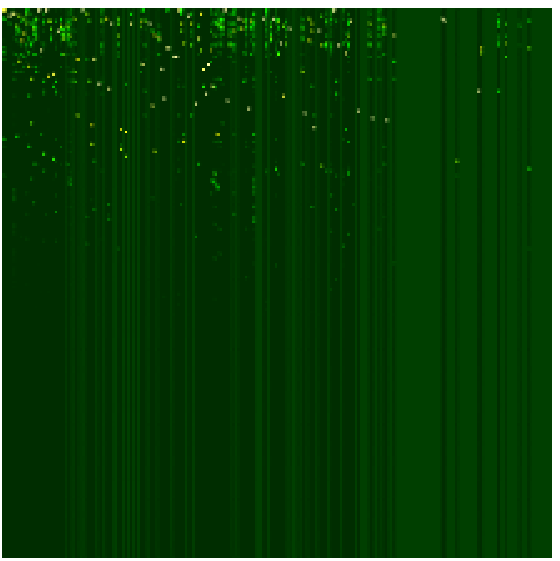} \\
\vspace{0.01\textwidth}
\includegraphics[width=.225\textwidth]{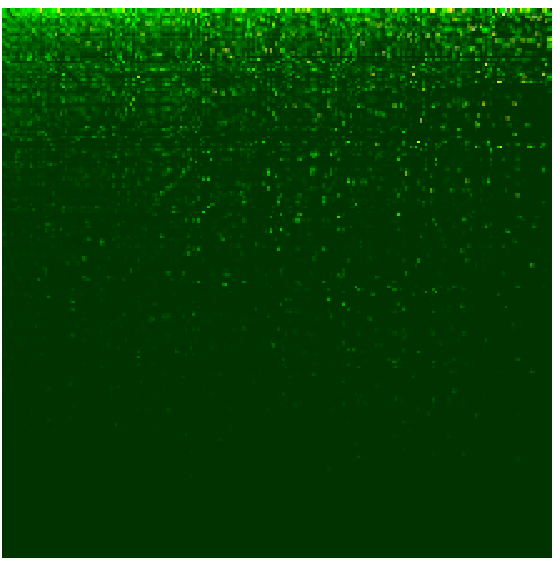} \hspace{0.01\textwidth}
\includegraphics[width=.225\textwidth]{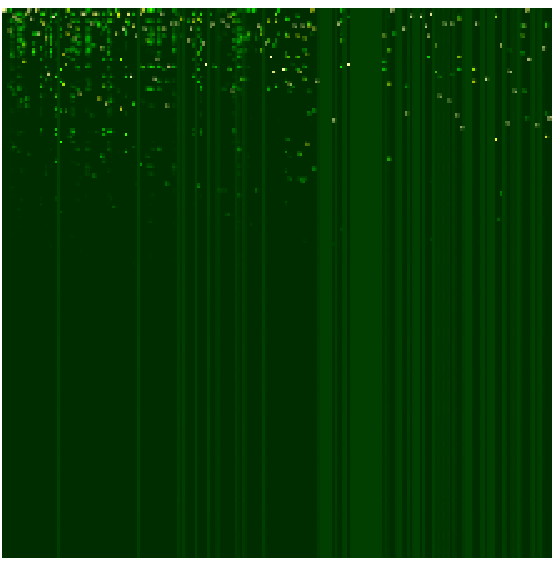} \\ 
\vspace{0.01\textwidth}
\includegraphics[width=.25\textwidth]{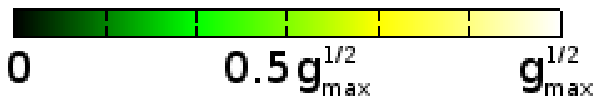} 
\caption{(Color online) Image of money mass matrix $M$ (top), 
Google matrix $G$ (middle) and inverse Google matrix $G^*$ (bottom) 
for \emph{all commodities} (left column) and \emph{crude petroleum} 
(right column) for year 2008 with all world countries $N=227$
from the UN COMTRADE \cite{comtrade}. 
Matrix elements $g$, for $M_{i,j}$, $G_{i,j}$ or $G^*_{i,j}$,
are shown by color changing from $0$ to a corresponding 
maximum value $g_{max}$.   All three matrices are shown in the basis of
PageRank index $K$ (and $K'$) of matrix $G$, respectively
for \emph{all commodities} (left) and \emph{crude petroleum} (right), 
which correspond to $x,y$-axis with $1 \leq K,K' \leq N$.
Here we use $\alpha=0.5$ for matrix $G$ and its PageRank index $K$ 
and the same $\alpha$ for $G^*$.  
}
\label{fig1}
\end{center}
\end{figure}

The PageRank performs ranking determined by ingoing links
putting at the top most known and popular nodes.
However, in certain networks outgoing links 
also play an important role.
Recently, on an example  of procedure call network of
Linux Kernel software, it was shown \cite{alik} that 
a relevant additional ranking is obtained by    
taking the network
with inverse link directions in the adjacency matrix
corresponding to $A_{ij} \rightarrow A^T=A_{ji}$
and constructing from it an additional Google matrix $G^*$
according to relation (\ref{eq1}) at the same $\alpha$. 
The examples of matrices $G$ and $G^*$
for the world trade network are shown in Fig.~\ref{fig1}.
The eigenvector of $G^*$ with eigenvalue 
$\lambda=1$ gives then a new inverse PageRank $P^*(i)$
with ranking index $K^*(i)$. This ranking was named 
CheiRank \cite{wiki} to mark that it allows to
{\it chercher l'information} in a new way.
While the PageRank rates the network nodes in average 
proportionally to a number of ingoing links, the CheiRank 
rates nodes in average proportionally to a number of outgoing links. 
The results obtained in \cite{alik,wiki}
confirm this proportionality with the exponent
$\beta_{out}=1/(\gamma_{out}-1)$.

Since each node belongs both to CheiRank and PageRank vectors 
the ranking of information flow on a directed network 
becomes {\bf two-dimensional}.
While PageRank highlights how  popular and known is
a given node, CheiRank highlights its communication 
and connectivity abilities.
The examples of Linux and Wikipedia networks
show that the rating of nodes based on PageRank and CheiRank
allows to perform information retrieval 
and to characterize network properties in a 
qualitatively new way \cite{alik,wiki}.

In this work we apply CheiRank and PageRank 
approach to the World Trade Network (WTN) using the 
enormous and detailed United Nations Commodity Trade
Statistics Database (UN COMTRADE) \cite{comtrade}.
Using these data we analyze the world trade
flows both in import and export for {\it all commodities}
for all years 1962 - 2009 available there
at SITC1 and HS96 databases. We also performed analysis
for specific commodities taken from SITC Rev. 1
data\-base, mainly for year 2008:
{\it crude petro\-leum} (S1-33101, "Crude petroleum"),
{\it natural gas} (S1-3411, "Gas, natural"), 
{\it barley} (S1-0430, "Barley, unmilled"),
{\it cars} (S1-7321, "Passenger motor cars, other than buses"),
{\it food} (S1-0, "Food and live animals"),
{\it cereals} (S1-04, "Cereals and cereal preparations").
Their codes and official UN names are given in brackets. 
In few cases, when certain 
countries were non-reporting their
export, we complemented the WTN data from the
import database.

For a given year we extract from the UN COMTRADE  
money transfer (in {\it USD}) from country
$j$ to country $i$ that gives us money matrix elements
$M_{ij}$ (for all types of commodities noted above).
These elements can be viewed as a money mass transfer from $j$ to $i$.
In contrast to the adjacency matrix $A_{ij}$ of WWW, where all elements
are only $0$ or $1$, here we have the case of weighted
elements. This corresponds to a case when there are
in principle multiple number of links from $j$ to $i$
and this number is proportional to {\it USD} amount transfer. 
Such a situation appears for rating of scientific journals
\cite{bergstrom}, Linux PCN \cite{alik} 
and for Wikipedia English articles 
hyperlink network \cite{wiki}, where generally
there are few citations (links) from 
a given article to another one. In this case still
the Google matrix is constructed according to the
usual rules and relation
(\ref{eq1}) with $S_{ij}=M_{ij}/m_j$
and $S_{ij}=1/N$, if for a given $j$ 
all elements $M_{ij}=0$.
Here $m_j= \sum_i M_{ij}$ is the total export mass
for country $j$. The matrix $G^*$
is constructed from transposed 
money matrix with $S_{ij}=M_{ji}/\sum_i M_{ji}$.
In this way we obtain the Google matrices $G$ and $G^*$
of WTN which allow to treat all countries on equal grounds 
independently of the fact if a given country is rich or poor.
A similar choice was used in rating of 
scientific journals \cite{bergstrom}, PCN Linux \cite{alik}
and Wikipedia network \cite{wiki}. The main difference
appearing for WTN is a very large variation of mass matrix elements
$M_{ij}$ related to the fact that there is very strong variation
of richness of various countries. Due to these reason we think that
it is important to use the ranking based on the Google matrix
which treats in a democratic way all world countries
that corresponds to the democratic standards of the UN.
For the WTN CheiRank and PageRank are naturally linked to 
export and import flows for a given country and hence it is 
very natural to use these ranks for characterization of country trade 
abilities. The Google matrix can be constructed in the same way 
not only for {\it all commodities}
but also for a given specific commodity. 

We note that recently the interest to the analysis of the world trade
as a network becomes more and more pronounced
with a few publications in this area 
\cite{garlaschelli,vespignani,benedictis,hedeem}. 
Thus, the global network characteristics
were considered in \cite{garlaschelli,vespignani},
degree centrality measures were analyzed in \cite{benedictis}
and time evolution of network global characteristics 
was studied in \cite{hedeem}. Topological and clustering
properties of multinetwork of
various commodities were discussed in \cite{garlaschelli2010}. 
Here we present a
systematic study of directed WTN on the basis of new combination
of PageRank and CheiRank methods using the Google matrix
constructed for 
the enormous UN COMTRADE database.

The paper is composed as follows: in Section 2 we
describe the global properties of the Google matrix of WTN,
in Section 3 we analyze distribution of countries
in PageRank-CheiRank plane for all time
period 1962 - 2009 and propose a random matrix model
of WTN (RMWTN)
which describes the statistical properties
of this distribution in the case of {\it all commodities};
comparison with ranking based on import and export
for various commodities is presented in Section 4;
discussion of the results is given in Section 5.
More detailed information and data are
given in Appendix and at the website \cite{cheirankpage}.

\section{Properties of Google matrix of WTN}
An example of the Google matrix of WTN in 2008
is shown in Fig.~\ref{fig1} for {\it all commodities} 
and {\it crude petroleum}. The matrices $G$ and $G^*$
are shown in the bases where all countries are ordered
by the PageRank index $K$ of matrix $G$ 
constructed for corresponding commodity 
(left and right columns). The matrix elements of $G$
are distributed over all $N$ values being
roughly homogeneously in $K$, even if the left top corner
at small $K$, $K'$ values is filled in a more dense way.
In contrast the density drops at large values of $K'$.
Such a structure is visible both for 
{\it all commodities} and {\it crude petroleum}
but clearly the global density is smaller in the later case 
since there are less number of links there
(see data in next Section).
The structure of $G^*$ is approximately the same
(we will see in next Section that
rich countries are located at low $K$, $K'$ values).
In contrast to $G$ and  $G^*$ the structure of money 
matrix $M$ is rather different.
For {\it all commodities} matrix elements
drop very rapidly at large values of $K$ and $K'$
that corresponds to the fact that the main amount of 
world money circulates only between rich countries
with top ranks $K$. In contrast to that
for {\it crude petroleum} the matrix elements
of $M$ are located at intermediate 
$K$ values. Indeed, in this case
PageRank index $K$ orders countries
by their  {\it crude petroleum} trade
where richest countries are not necessarily
at the top ranks.

\begin{figure}[ht]
\begin{center}  
\includegraphics[width=.45\textwidth]{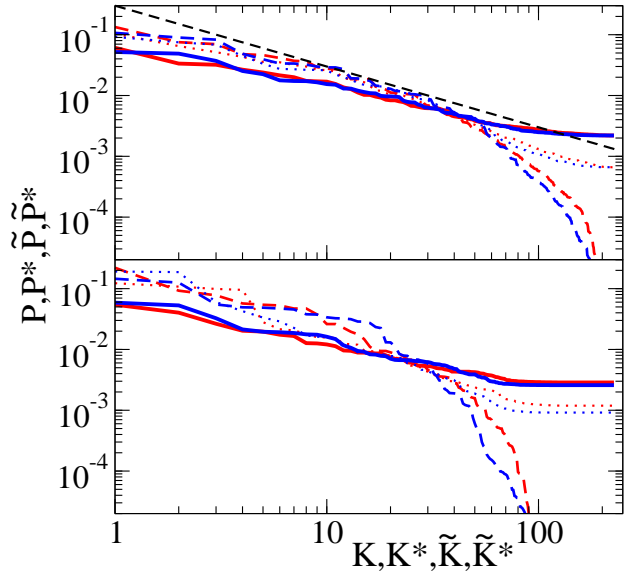}
\caption{(Color online)
Probability distributions of PageRank $P(K)$, CheiRank $P^*(K^*)$, 
ImportRank $\tilde{P}(\tilde{K})$, and ExportRank $\tilde{P^*}(\tilde{K^*})$
are shown as function of their indexes in  logarithmic scale 
for \emph{all commodities} (top panel) and \emph{crude petroleum} 
(bottom panel) for WTN in 2008 with $N=227$.
Here $P(K)$ and $P^*(K^*)$ are shown by red and blue 
curves respectively, 
for $\alpha=0.5$ (solid curves) and $\alpha=0.85$ (dotted curves);
$\tilde{P}(\tilde{K})$ and $\tilde{P}^*(\tilde{K}^*)$ 
are displayed by dashed red and blue curves respectively.
For both commodities the distributions
$P(K)$ and $P^*(K^*)$  
follow a power law dependence like $P \propto 1/K^{\beta}$
(see text), the Zipf law is shown by the straight 
dashed line with $\beta=1$ in top panel.
}
\label{fig2}
\end{center}
\end{figure}

From the Google matrices $G$ and $G^*$
we find the probability distributions PageRank $P(K)$
and CheiRank  $P^*(K^*)$ which are shown in Fig.~\ref{fig2}
for the same commodities as Fig.~\ref{fig1}.
One of the main features of these distributions 
is that both $P(K)$ and $P^*(K^*)$ depend
on their  indexes
in a rather similar way from,
that is in contrast to the results
found for the WWW \cite{donato,upfal}, PCN Linux \cite{alik}
and Wikipedia network \cite{wiki},
where these distributions are different
having different exponents
$\beta$ in the power law decay.
Here, up to fluctuations, we have 
$\beta_{in} = \beta_{out}=\beta$.
The size of WTN is rather small
compared to usual sizes of WWW,
Linux or Wikipedia networks.
However, still we find that the power law
gives a quite good fit of our data.
The fit gives $\beta=1.17 \pm 0.015$ at
$\alpha=0.85$ and  $\beta=0.63 \pm 0.01$ at $\alpha=0.5$
(for {\it all commodities})
 and $\beta = 0.92 \pm 0.02$ and $\beta=0.51 \pm 0.01$
respectively (for {\it crude petroleum})
for all  227 countries in Fig.~\ref{fig2}.
For the fit of top 100 countries
we have respectively
$\beta=1.15 \pm 0.03$ ($\alpha=0.85$) and
$\beta=0.75 \pm 0.008$ ($\alpha=0.5$) for {\it all commodities}
and $\beta=1.22 \pm 0.015$ ($\alpha=0.85$) and
$\beta=0.70 \pm 0.008$ ($\alpha=0.5$) for {\it crude petroleum}.
There is a certain change of the exponent
with a decrease of the fit interval
which, however, is not very large. We attribute this 
to visible deviations 
at the tail of $K$, $K^*$ with small 
countries (see discussion in next Section).  In average
the exponent value is not very far from the
value $\beta=1$ corresponding to the Zipf law
\cite{zipf}. 
\begin{figure}[ht]
\begin{center}  
\includegraphics[width=.45\textwidth]{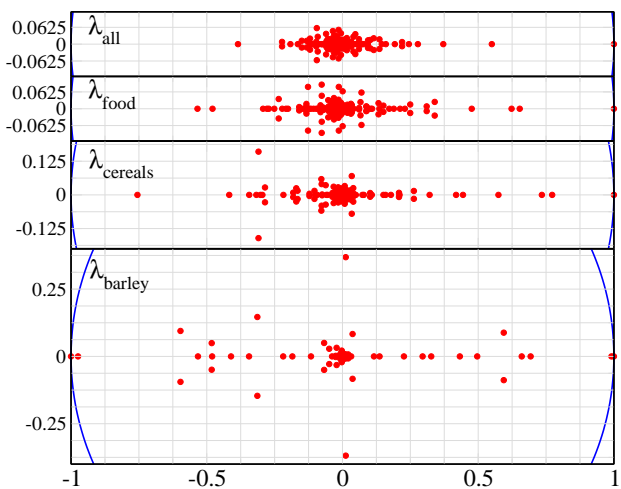}
\caption{(Color online)
Spectrum of the eigenvalues $\lambda$ of the
Google matrix at $\alpha=1$ in complex plane 
for WTN in 2008 with $N=227$ countries for
\emph{all commodities}, \emph{food}, \emph{cereals} and \emph{barley} 
(from top to bottom); all eigenvalues are shown for
each commodity; unit radius circle is also shown.
Only the bottom case have quasi-degenerate eigenvalues
close to the circle with $3$  values
$\lambda=1., 0.99987, 0.991$  and two values close to $\lambda=-1$;
other cases have a significant gap separation from $|\lambda|=1$.
}
\label{fig3}
\end{center}
\end{figure}

It is useful to compare the behavior of probabilities
$P(K)$ and $P^*(K^*)$ with 
respective ranking related to
import and export. To do that
we rank the countries by
probability import $\tilde{P}(\tilde{K})$
defined as a ratio of import in {\it USD}
for a given country $\tilde{K}$
to the total world import in {\it USD} for a given year
with ordering of all countries in decreasing 
probability order index of ImportRank $\tilde{K}$.
By construction we have $\sum_{\tilde{K}}\tilde{P}(\tilde{K})=1$
and $\tilde{P}(\tilde{K})=m_{\tilde{K}}/M_T$,
where $m_{\tilde{K}}$ is the import mass of a given country 
$\tilde{K}$ and $M_T = \sum_{i,j} M_{ij}$
is the total world money mass for a given year.
In the same way we construct export probability
$\tilde{P^*}(\tilde{K^*})$ with the ExportRank $\tilde{K^*}$. 
The dependence
of these probabilities on their indexes is shown
in Fig.~\ref{fig2}. In the range of
$1 \leq \tilde{K}, \tilde{K^*} \leq 50$
it can be well described by a power law with
$\beta = 1.01 \pm 0.03$ 
for {\it all commodities}
corresponding to the Zips law
(for {\it crude petroleum} we obtain for this range
$\beta = 1.43 \pm 0.07$).
At larger values of order index
we find a sharp drop with
an exponential type decay on the tail.
For {\it crude petroleum} this exponential
decay starts at smaller values of $K$
due to a significantly smaller 
total number of links that 
gives an increase of $\beta$
for the range of 50 countries. 
The exponential decay
at large $K$ results from
a strong variation of richness of countries
which changes more than by four orders of magnitude.
From the comparison of ranks shown in Fig.~\ref{fig2}
it is clear that
PageRank and CheiRank give more 
equilibrated and democratic description of
trade flows.

We should note that due to a small size of the WTN
the fluctuations are stronger compared to large size 
networks like the WWW. It is especially visible
for specific commodities where the total
number of links is by factor 30 smaller then for
{\it all commodities} (see next Section).
These fluctuations are smaller for 
the damping factor value $\alpha=0.5$
in agreement with the results presented in
\cite{avrachenkov,ggs}. In fact this $\alpha$
value was also used in \cite{redner}
for PhysRev citation network. Due to that reasons
in next Sections we show data for 
ranking at $\alpha=0.5$.

Finally let us discuss the spectrum $\lambda$
of the Google matrix which follows from the equation for 
right eigenvectors $\psi_m(i)$:
\begin{equation}
  \sum_j G_{ij} \psi_m(j) = \lambda_m \psi_m(i) \;\; .
\label{eq2} 
\end{equation} 
It is known that the dependence on $\alpha$ is rather simple:
all eigenvalues, except one with $\lambda=1$,
are multiplied by $\alpha$ \cite{googlebook}.
Due to that we show the spectrum of $G$ at $\alpha=1$
in Fig.~\ref{fig3}. Compared to the spectrum 
studied for other networks 
(see examples in \cite{alik,ggs,ulam1,ulam2,ulam3,weylinux})
we find that the WTN spectrum is very close to real line
especially for three top commodities in Fig.~\ref{fig3}.
We explain this by the fact that here an average
number of links per country is very large
for these commodities and that the matrix elements
are not very far from the symmetric relation
$M_{ij} = M_{ji}$ at which the spectrum is real.
We only note that  for  {\it barley}
the spectrum has quasi-degeneracy at
$\lambda=1$ that signifies the existence of slow
relaxation modes. We attribute this to the fact that
there are certain countries which practically
do not use {\it barley} that leads to appearing of
isolated subspaces with
corresponding quasi-degenerate modes.
We will return to the discussion 
of spectrum properties of $G$ in next Section.

\section{CheiRank versus PageRank for WTN}
\begin{figure}[h!]
\begin{center}  
\includegraphics[width=.4\textwidth]{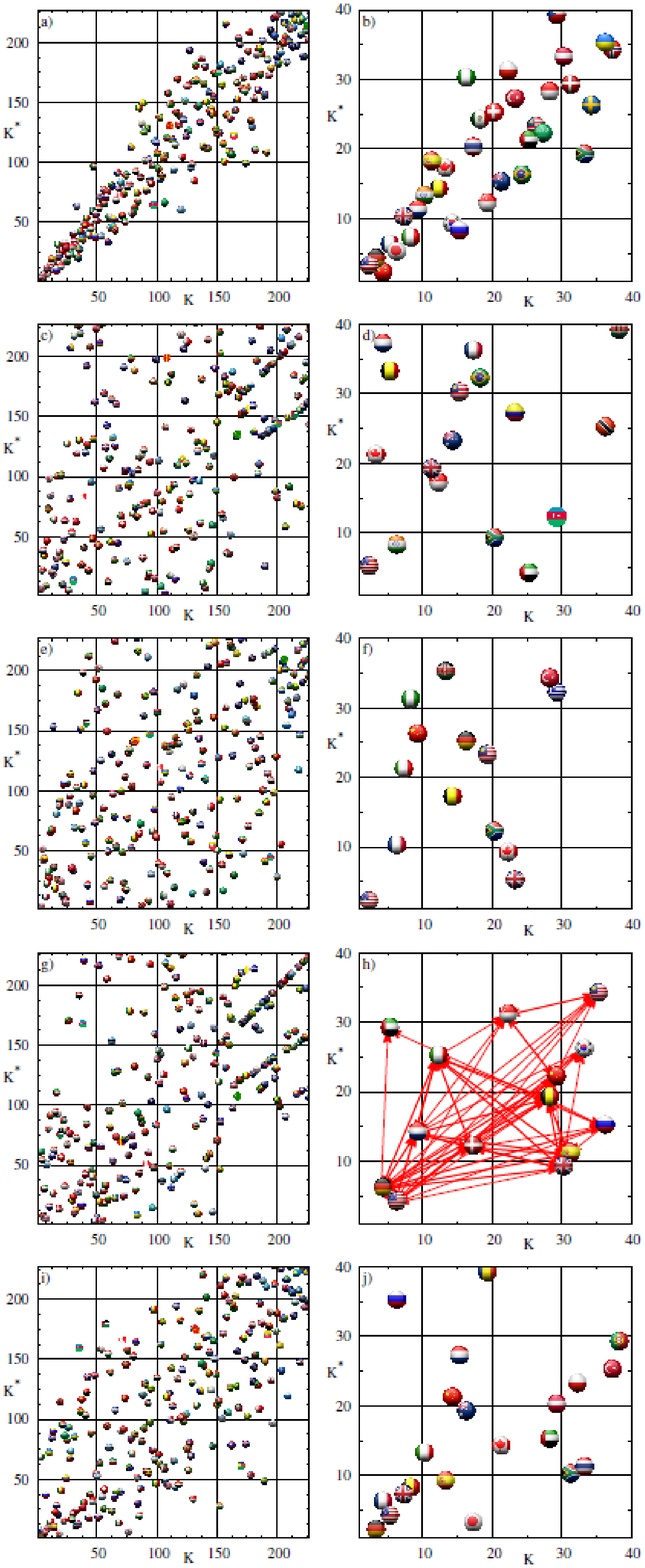}
\caption{(Color online)
Country positions in PageRank-CheiRank plane $(K,K^*)$ 
for world trade in various commodities in 2008.
Each country is shown by circle with its own flag
(for a better visibility
the circle center is slightly displaced from its
integer position $(K,K^*)$ along direction angle $\pi/4$).
The panels show the ranking for trade in the
following commodities: \emph{all commodities} $(a,b)$;
\emph{crude petroleum} $(c,d)$;
\emph{natural gas} $(e,f)$;
\emph{barley} $(g,h)$; 
\emph{cars} $(i,j)$.
Left column shows a global scale with all 227 countries, 
while right column gives a zoom in  the region of $40\times40$ top ranks.
For \emph{barley} in panel (h) the links between countries 
inside the selected region are shown by arrows.
}
\label{fig4}
\end{center}
\end{figure}

We start from examples of distributions of countries
in the PageRank-CheiRank plane shown in Fig.~\ref{fig4}
for 5 different commodities in 2008. 
The first case of 
{\it all commodities} corresponds to
trade flows between countries integrated over all 
type of products. Even if the Google matrix approach
is based on a democratic ranking of
international trade, being independent of
total amount of export-import for a given 
country, we still find at the top ranks $K$ and $K^*$
the group of industrially developed countries
(see more details in Table 1 in Appendix).
This means that these countries have
efficient trade networks with optimally
distributed trade flows. Another pronounced feature 
of global distribution is that it is
concentrated along the main diagonal $K=K^*$.
This feature is not present in 
other networks studied before (e.g. PCN Linux \cite{alik}
and Wikipedia \cite{wiki}). The origin of this 
density concentration is based on simple economy
reason: for each country the total import  is 
approximately equal to export since each country should keep
in average an economic balance.
Thus for a given country its trade is doing well
if its $K^*< K$ so that the country
exports more than it imports. The opposite
relation $K^*> K$ corresponds to a bad trade situation.
We also can say that local minima
in the curve of $(K^*-K) \; vs. \; K$ correspond to 
a successful trade while maxima mark
bad traders. In 2008 most successful were
China, Rep. of Korea, Russia, Singapore,
Brasil, South Africa, Venezuela (in order of
$K$ for $K \leq 50$) while among bad traders we note
UK, Spain, Nigeria, Poland, Czech Rep., Greece, Sudan
with especially strong export drop for two last cases.
The comparison of our ranking with the import-export
ranking will be analyzed in next Section.

Even if there is a concentration of density along the main
diagonal (Fig.~\ref{fig4}a) we  still have
a significant broadening of distribution especially
at middle values of $K \sim 100$.
This means that the gravity model of trade, often used in 
economy (see e.g. \cite{krugman2011,benedictis}),
has only approximate validity. Indeed, in this 
model the mass matrix $M_{ij}$ is symmetric
that would placed all countries on
diagonal $K=K^*$ that is definitely not the case.
\begin{figure}[ht]
\begin{center}  
\includegraphics[width=.45\textwidth]{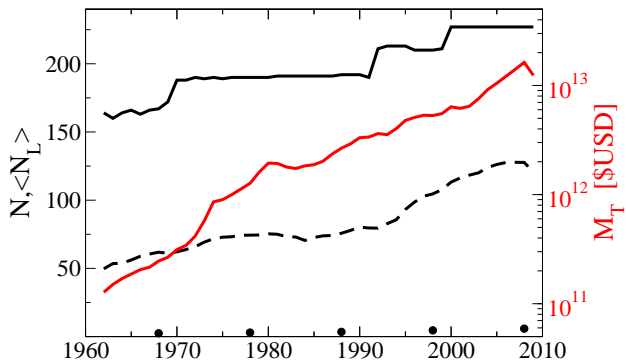}
\caption{(Color online)
Time evolution of the number of countries ($N$, full black curve), 
average number of links per country ($\langle N_L\rangle$)
for  \emph{all commodities} (dashed curve) and
\emph{crude petroleum} (points for five years),
total amount of money ($M_T =\sum_{i,j} M_{ij}$, red curve).
The scale of $N$ and $\langle N_L\rangle$ is shown on left side, 
while $M_T$ values,  in $\${\it USD}$, 
are given in logarithmic scale on the right side.
}
\label{fig5}
\end{center}
\end{figure}

If we now turn to the distribution of countries
for a trade in a specific commodity
than it becomes absolutely clear that
the symmetry approximately visible for {\it all commodities}
is absolutely absent: the points are scattered 
practically over the whole square $N \times N$.
The reason of such a strong scattering is clear:
e.g. for  {\it crude petroleum} some countries 
export this product while other countries 
import it. Even if there is some flow from exporters
to exporters it remains relatively low
(see more discussion in next Section).
This makes the Google matrix to be very asymmetric.
Indeed, the asymmetry of trade flow is well visible in
Fig.~\ref{fig4}h where arrows show the trade directions
between countries within top $40 \times 40$ ranks
for {\it barley}.

It is also useful to use 2DRank $K_2$ discussed in \cite{wiki},
which orders all nodes according to the order of their
appearance inside squares of  size $K \times K$
going from $K=1,2,3,...$ to $N$
for each of four specific commodities shown in Fig.~\ref{fig4}.
In a certain sense top countries in 2DRank $K_2$
are those which are active traders even not being among large
exporters or importers of this product
(all ranks for commodities of Fig.~\ref{fig4} are given in
Tables 1-5 in Appendix). As an example, we note
Singapore which is at the third position in $K_2$
(Table 2): it is a small country which
cannot export or import a large amount of the commodity,
but its trade network is very active redistributing
flows between various countries that
places it at a high $K_2$ rank.

The images of Fig.~\ref{fig4} allow to understand qualitatively
the reasons of density concentration around diagonal 
$K=K^*$ for the case of {\it all commodities}:
this trade is composed from hundreds of specific commodities which
behave  randomly and the averaging over them 
gives effective coarse-graining and
produces a certain symmetry for matrix elements
due to the central limit theorem for
a sum of many positive contributions.
The fact that the increase of coarse-graining cell
gives more and more symmetry is well seen in Fig.~\ref{fig3}
where the spectrum becomes more and more close to a real one,
and hence there is more and more symmetry in elements $G_{ij}$,
when we go from {\it barley} to 
{\it cereals}, {\it food} and {\it all commodities}.

We will return to the analysis of specific country ranking 
in the next Section while now we turn to analysis
of time evolution of WTN.
\begin{figure}[h!]
\begin{center}  
\includegraphics[width=.45\textwidth]{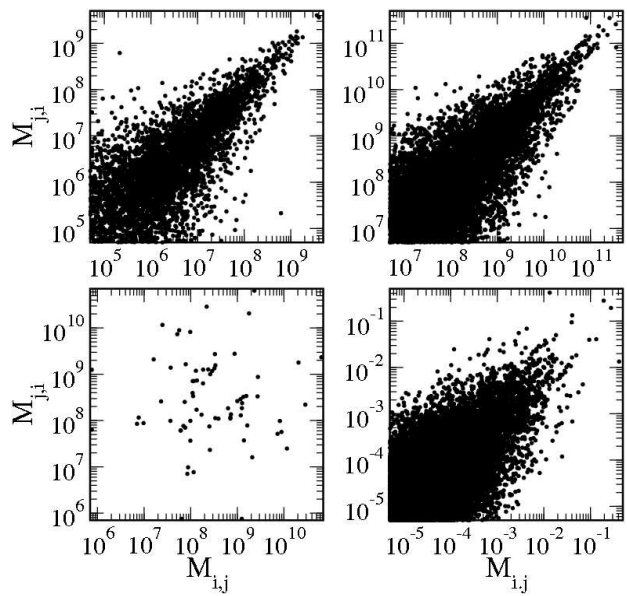}
\caption{
Money mass transfer matrix elements $M_{i,j}$ 
are shown versus their transposed values $M_{j,i}$ 
for \emph{all commodities} of WTN in 1962 (top left panel)
and 2008 (top right panel).
Bottom left panel shows the matrix elements
for \emph{crude petroleum} of WTN in 2008; 
bottom right panel shows the same quantities for random matrix model of WTN.
Four panels show 5 orders of magnitude in logarithmic scales
starting from maximum values of $M_{ij}$. 
In the case of WTN (top and bottom left panels) 
matrix elements are taken from the  UN COMTRADE database 
and are expressed in {\it USD},
right bottom panel is built from one random realization
with $M_{ij}= \epsilon_i \epsilon_j/ij$ (see text).
Here $N=164$ for 1962 data; $N=227$ for 2008 data
and RMWTN. 
}
\label{fig6}
\end{center}
\end{figure}

The variation of global parameters of WTN during the 
database period 1962 - 2009 is shown in Fig.~\ref{fig5}.
The number of countries is increased by 38\%,
while the number of links per country
for {\it all commodities}
is increased in total by  140\% with a significant increase
from 50\% to 140\%  
during the period 1993 - 2009
corresponding to economy globalization.
At the same time for a specific commodity
the average number of links per country
remains on a level of 3-5 links
being by a factor 30 smaller 
compared to {\it all commodities} trade.
During the whole period the total amount
$M_T$ of trade in {\it USD}
shows an average
exponential growth by 2 orders of magnitude.
\begin{figure}[ht]
\begin{center}  
\includegraphics[width=.45\textwidth]{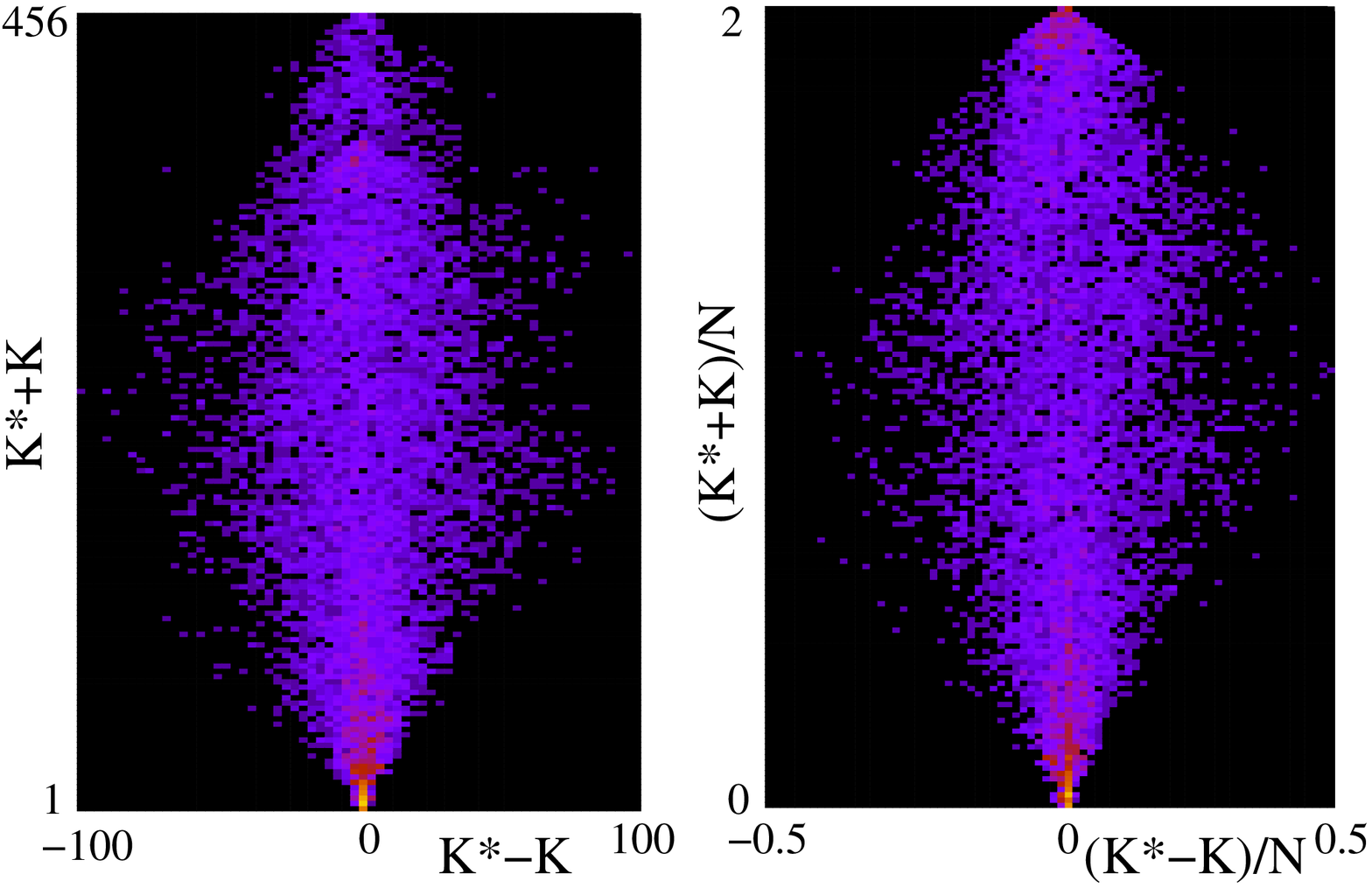}\\
\includegraphics[width=.45\textwidth]{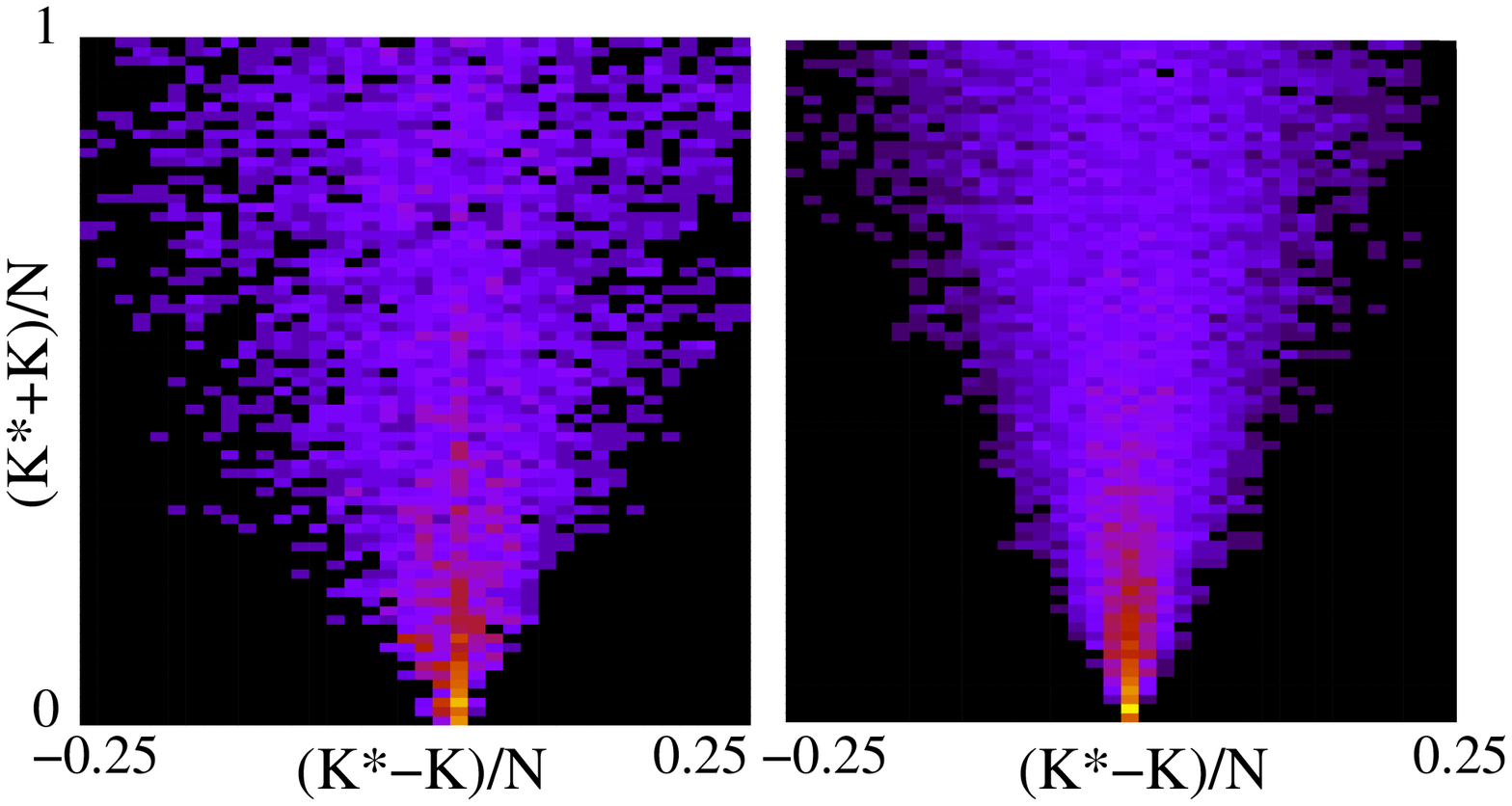}
\caption{(Color online) Spindle distribution for WTN
of \emph{all commodities} for all countries
in the period 1962 - 2009 shown in the plane of 
$(K^*-K, K^*+K)$ (left top panel, coarse-grained data 
in $3 \times 3$ cell size)
and in the rescaled plane $((K^*-K)/N, (K^*+K)/N)$
(right top panel, coarse-graining inside
each of $76 \times 152$ cells, which is approximately the same 
number as in top left panel); 
data from the UN COMTRADE database.
Bottom left panel: zoom of top right panel;
bottom right panel:  data from 100 realisations of RMWTN model
(\ref{eq3}) with  $N=227$ as for WTN size in 2008.
}
\label{fig7}
\end{center}
\end{figure}

To understand the physical properties of the WTN
we consider the distribution of money mass transfer
matrix elements $M_{ij}$ shown versus their 
transposed values  $M_{ji}$ in Fig.~\ref{fig6}.
This distribution is symmetric by the construction.
In the case of symmetric matrix $M_{ij}$,
corresponding to the gravity model of trade or undirected network,
all elements should be located on one diagonal 
line that is definitely not the case.
For {\it crude petroleum} the distribution 
is even more broad showing definite absence of symmetry
of $M_{ij}$.
In fact for {\it all commodities}
the distribution forms a rather broad cone
which form remains stable in time
according to the comparison of data
in 1962 and 2008 years (the density is higher 
in the later case since there are more countries).
Keeping in mind that according to data of 
Fig.~\ref{fig2} we have
the Zipf law for $\tilde{P}(\tilde{K})$
we propose the random matrix model of WTN (RMWTN)
with the mass matrix elements given by 
\begin{equation}
  M_{ij} = \epsilon_i \epsilon_j/ij \;\; ,
\label{eq3} 
\end{equation}
where $\epsilon_i$ are random numbers homogeneously 
distributed in $[0,1]$ interval and $i,j$
are indexes in the ImportRank index $\tilde{K}$.
The distribution given by this simple model
reproduces quite well the actual distribution
found for {\it all commodities}
(see right panels in Fig.~\ref{fig6}).
With this RMWTN distribution of $M_{ij}$
we construct the Google matrices $G$ and $G^*$ 
according to the usual recipes (\ref{eq1})
and then determine the distribution of points 
in $(K,K^*)$ plane.
\begin{figure}[ht]
\begin{center}  
\includegraphics[width=.45\textwidth]{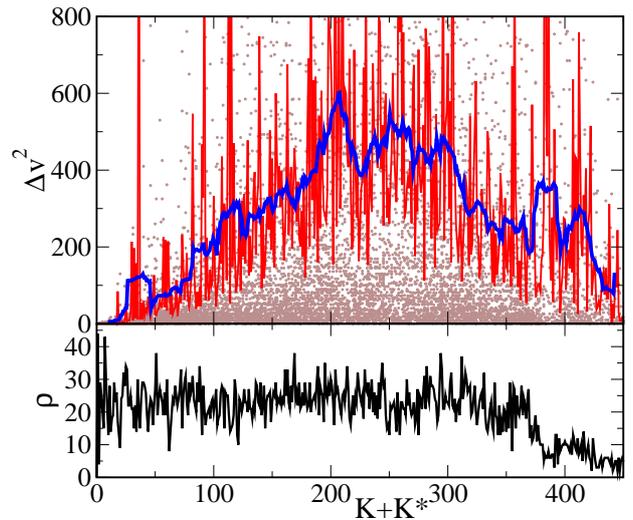}
\caption{(Color online)
Top panel shows velocity square $\Delta v^2$ 
as a function of $K+K^*$
for all countries and all years
({\it all commodities} data). 
Gray circles represent all values of $\Delta v^2$,
red curve shows the value of $\Delta v^2$ 
averaged over cases with fixed $K+K^*$, 
blue curve shows
the average of the red curve data in the interval $[K+K^*-10,K+K^*+10]$.
In the bottom panel the number of cases $\rho(K+K^*)$ 
at a given $K+K^*$ is shown 
as a function of $K+K^*$.
}
\label{fig8}
\end{center}
\end{figure}

To have a statistical comparison between the RMWTN
and real WTN data we construct the density
distribution of countries in the plane $(K^*-K, K^*+K)$
using {\it all available years 1962 - 2009
at the UN COMTRADE database} for  {\it all commodities}.
The coarse-grained distribution of about $10^4$
WTN data points is shown in Fig.~\ref{fig7}.
We present the data directly in $(K^*-K, K^*+K)$ 
plane (top left panel)
and in rescaled variables $((K^*-K)/N, (K^*+K)/N)$ plane,
which takes into account that the number of countries 
grown by 38\% during this time period.
The distribution has a form of {\it spindle}
with maximum density
at the vertical axis $K^*-K=0$.
We remind that good exporters are
on the left side of this axis at
$K^*-K < 0$, while the good importers (bad exporters)
are on the right side at $K^*-K > 0$.

\begin{figure}[ht]
\begin{center}  
\includegraphics[width=.45\textwidth]{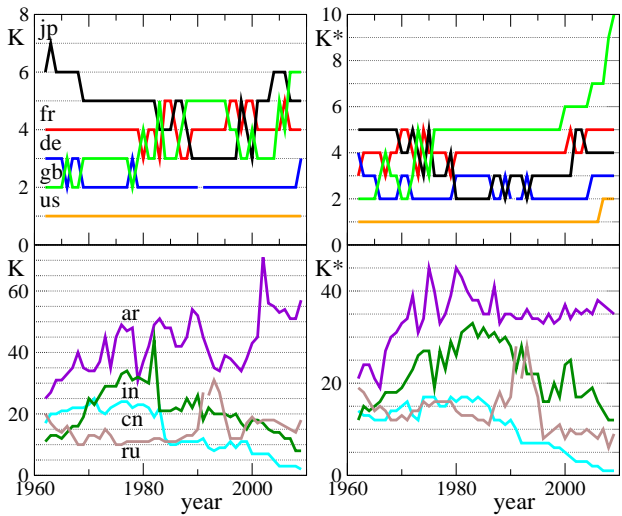}
\caption{(Color online) Time evolution of 
CheiRank and PageRank indexes $K$, $K^*$
for some selected countries for \emph{all commodities}. 
The countries shown in top panels are: 
Japan (jp-black), France (fr-red), Fed. Rep. of Germany 
and Germany (de - both in blue), Great Britain (gb - green), 
USA (us - orange) [curves
from top to bottom in 1962 in left top panel].
The countries shown in bottom panels are: 
Argentina (ar - violet), India (in - dark green), China (cn - cyan), 
USSR and Russian Fed. (ru - both in gray)
[curves from top to bottom in 1975 in left bottom panel].
}
\label{fig9}
\end{center}
\end{figure}

\begin{figure}[ht]
\begin{center}  
\includegraphics[width=.45\textwidth]{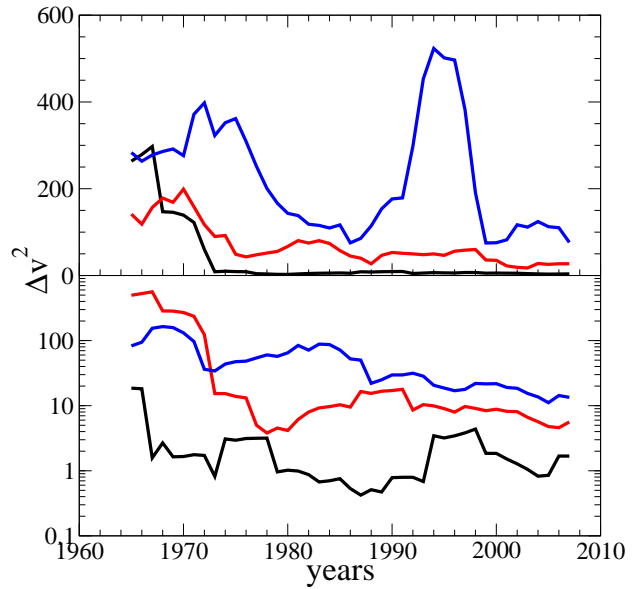}
\caption{(Color online) Time evolution of velocity square $\Delta v^2$
for {\it all commodities}
averaged over five years interval.
In addition $\Delta v^2$ is averaged over
countries in the following intervals: 
$1 \leq K+K^* \leq 40$ (blue curve),
$41 \leq K+K^* \leq 80$ (red curve),
$81 \leq K+K^* \leq 120$ (black curve) in top panel;
$1 \leq K+K^* \leq 20$ (blue curve),
$21 \leq K+K^* \leq 40$ (red curve),
$41 \leq K+K^* \leq 60$ (black curve) in bottom panel.
}
\label{fig10}
\end{center}
\end{figure}

The comparison of WTN data with the results
produced by RMWTN model (\ref{eq3}) are shown in 
bottom panels of Fig.~\ref{fig7}:
there is a good agreement between both without any
fit parameters for the half of all countries with
top ranks ($K+K^*<N$). For countries with
$K+K^*>N$ the RMWTN model does not succeed 
to describe correctly the upper part of spindle distribution 
found for the WTN and hence further improvements of the RMWTN 
are needed. However, a simple description of 
the distribution for a half top countries is
rather successful. 

A remarkable feature of the WTN spindle distribution
of Fig.~\ref{fig7} (top right) is the appearance of
high density domains at $K^*-K \approx 0$ with
$K+ K^* \approx 1$ and $K+ K^* \approx 2N$.
They give an impression of two solid phases 
emerging in these two regions while the other
part looks like a gas phase. This view gets
additional confirmation by data of Fig.~\ref{fig8}
where we present the velocity square $(\Delta v)^2$,
averaged over the whole period 1962 - 2009,
as a function of $K+ K^*$. This local quantity 
is defined as
$(\Delta v)^2=(K(t1)-K(t-1))^2+(K^*(t)-K^*(t-1))^2$
via a one year displacement of a given country 
in $(K,K^*)$ plane with further averaging
over all times and all countries.  
These data clearly show
that for $K+ K^* \leq 20$ we have very small
square velocity (small effective temperature)
corresponding to a solid phase of rich countries,
while for $K+ K^* > 20$ we have 
large square velocity (large effective temperature)
corresponding to a gas phase with rapid 
rank fluctuations. There is a similar visible drop
of temperature at another limit 
of most poor countries with $K+ K^* \approx 2N$
which indicates a formation of solid phase of
poor countries (the data are not so exact for
this region due to variation of number of UN countries with time).

\begin{figure}[ht]
\begin{center}  
\includegraphics[width=.45\textwidth]{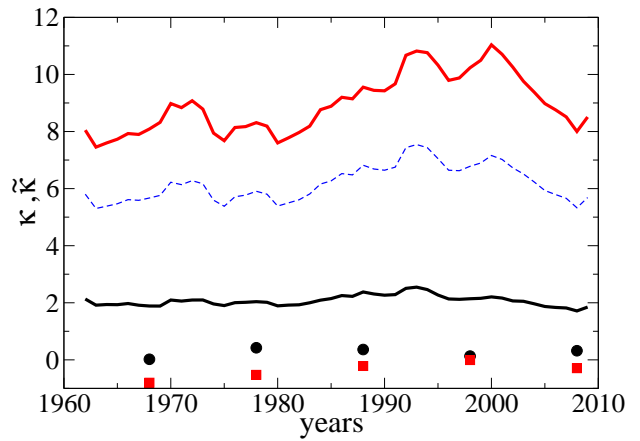}
\caption{(Color online)
Time evolution of correlators of 
PageRank--CheiRank ($\kappa$) and ImportRank--ExportRank ($\tilde{\kappa}$).
\emph{All commodities} are shown by solid red curve 
for $\tilde{\kappa}$, and solid black curve and dashed blue 
curve for $\kappa$ with $\alpha=0.5$ and $\alpha=0.85$ respectively.
Correlators for \emph{crude petroleum} 
with 10 years of separation are shown in 
red squares for $\tilde{\kappa}$ and black circles for $\kappa$.
}
\label{fig11}
\end{center}
\end{figure}

The presence of solid phase of rich countries
and gas phase of other countries
is also visible from analysis of rank variation in time
for individual countries shown in Fig.~\ref{fig9}:
for $K,K^* \leq 10$ the curves are almost flat
while for  $K,K^* > 10$ we see strong fluctuation
of curves. 
It is interesting to note that  sharp increases
in $K$ mark crises in 1991, 1998 for Russia 
and in 2001 for Argentina 
(import is reduced in period of crises).
We also see that in recent years the solid phase
is perturbed by entrance of new countries like China and India.
However, the results presented in Fig.~\ref{fig10} for
the variation of square velocity with time
for three regions of $K+ K^*$ show that the top 10,
and even top 20, countries 
have rather small velocities $\Delta v^2$,
compared to those with $(K+K^*)/2 \approx K > 20$.
For $K \leq 20$ we have $\Delta v^2$
which remains constant in time.
In a certain sense it looks that the 
countries with $20<K<40$ protect
those with $1 \leq K \leq 20$
(approximately corresponding to {\it G-20} 
major economies \cite{g20}), 
so that their temperature at $1 \leq K \leq 20$
remains unaffected even by a very larger
fluctuation well visible
for the range $81 \leq K+K^* \leq 120$
during the period of 1992 - 1998
with a few financial crises of 
Black Wednesday, Mexico crisis,
Asian crisis and Russian crisis.

The presented results for distribution of countries
and analysis of their time evolution in the PageRank-CheiRank plane
confirm a well known statement that
{\it ``the poor stay poor and the rich stay rich''}.

Finally let us discuss an additional
parameter which characterizes the correlation
between PageRank and Chei\-Rank vectors.
The correlator between PageRank and Chei\-Rank is defined as
\begin{equation}
  \kappa=N\sum_i P(K(i)) P^*(K^*(i)) -1 \; ,
\label{eq4} 
\end{equation}
and in a similar way the correlator between
ImportRank and ExportRank is given by
\begin{equation}
\tilde{\kappa}=N\sum_i \tilde{P}(\tilde{K}(i)) 
\tilde{P}^*(\tilde{K}^*(i)) -1 \; .
\label{eq5} 
\end{equation}
Recently it has been found that  
there are networks with small correlator,
like PCN Linux \cite{alik},
and large correlator, as Wikipedia \cite{wiki}.
Here we find that for {\it all commodities}
we have large values of $\kappa$ 
and $\tilde{\kappa}$, which have rather similar dependence on time
(see Fig.~\ref{fig11}). In contrast, there are
almost zero or even negative correlations
for {\it crude petroleum}. Indeed, for {\it crude petroleum}
there is no correlation between export and import
while for {\it all commodities}
they are strongly correlated.

\section{Comparison with Import - Export ranking}
It is important to compare rating based on PageRank and CheiRank
with the useful way of country rating based on
ImportRank and ExportRank (see e.g \cite{cia2010}).
With this aim we present the distribution of
country positions on the planes
$(\tilde{K},K)$ and $(\tilde{K^*},K^*)$
shown for top 100 for the same commodities
as in Fig.~\ref{fig4} for year 2008. For {\it all commodities}
there is a clear correlation between PageRank and ImportRank
since the distribution of points is centered along the diagonal
$K=\tilde{K}$. It starts to spread only around 
$K \approx \tilde{K} \approx 30$. At the same time
for CheiRank and ExportRank such a spreading from diagonal 
starts significantly earlier at $\tilde{K^*} \approx K^* \approx 10$.

\begin{figure}[h!]
\begin{center}  
\includegraphics[width=.4\textwidth]{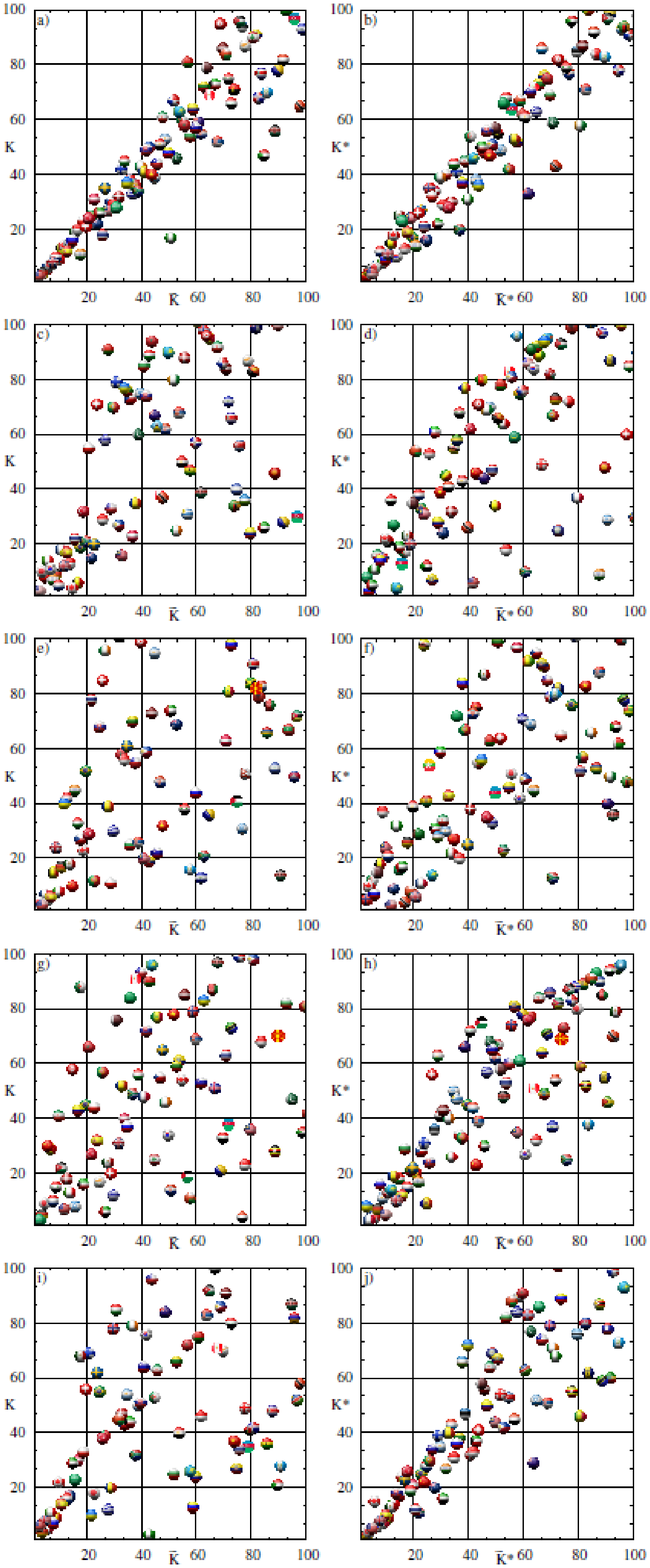}
\caption{(Color online)
Comparison of ranking between PageRank $K$ and 
ImportRank $\tilde{K}$ (left column), 
and between CheiRank $K^*$ and 
ExportRank $\tilde{K}^*$ (right column) for year 2008.   
The shown commodities are: 
\emph{all commodities} (panels a, b);
\emph{crude petroleum} (panels c, d);
\emph{natural gas} (panels e, f);
\emph{barley} (panels  g, h); 
\emph{cars} (panels i, j). Only top 100 ranks are shown.
}
\label{fig12}
\end{center}
\end{figure}

For other commodities shown in Fig.~\ref{fig12}
the correlations between ranking based on Google matrix and
corresponding Export or/and Import ranking
are practically absent showing very broad scattering of points
around almost the whole plane. Only for {\it cars}
there is a certain level of correlation for approximately 
the first 10 ranks. Natural products  like {\it crude petroleum},
{\it natural gas} and  agriculture products like 
{\it barley} show no correlations.

\begin{figure}[h!]
\begin{center}  
\includegraphics[width=.4\textwidth]{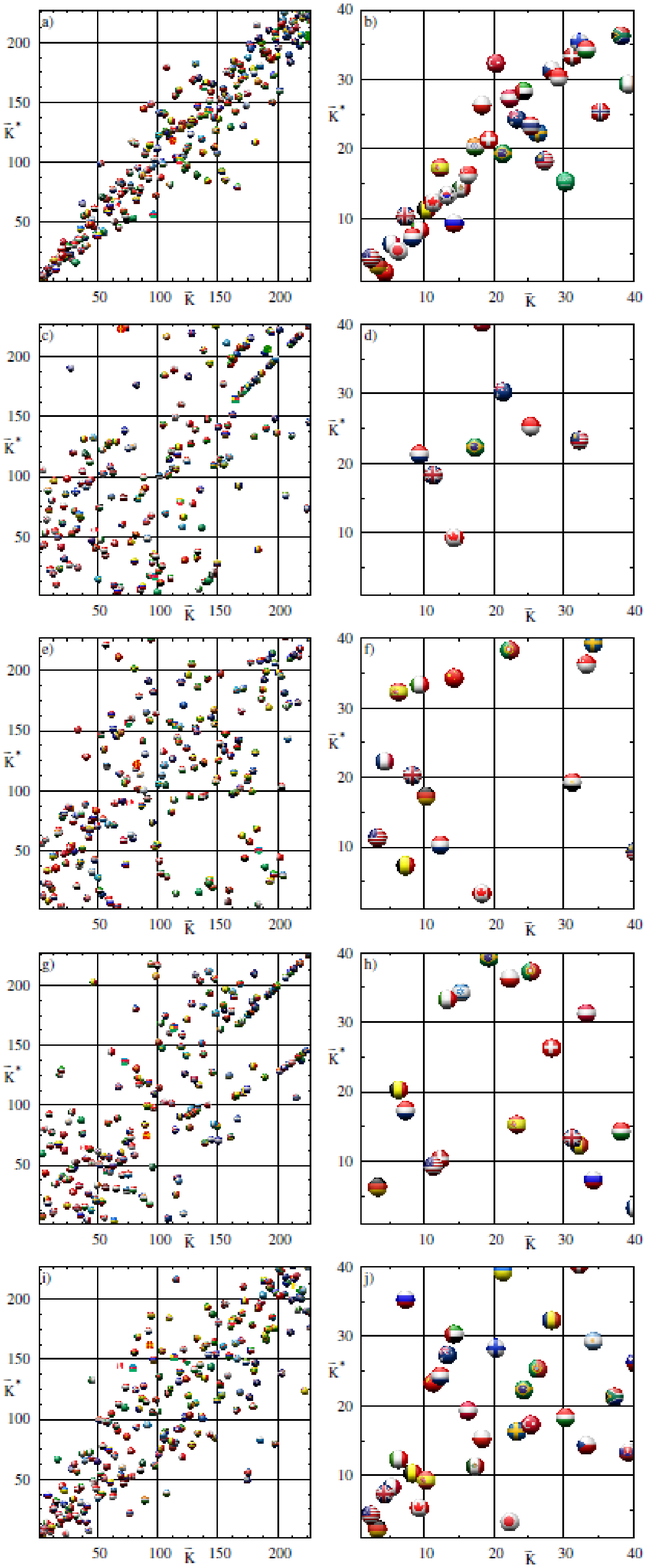}
\caption{(Color online)
Country positions in the ImportRank-ExportRank 
plane $(\tilde{K},\tilde{K}^*)$ for year 2008.
The shown commodities are: 
\emph{all commodities}  (panels a, b);
\emph{crude petroleum} (panel c, d);
\emph{natural gas} (panel e, f);
\emph{barley} (panel g, h); 
\emph{cars} in i) and j).
Left column shows a global scale (227 countries) 
while right column illustrates the first $40\times40$ region.
Data can be compared with those in Fig.~\ref{fig4}.
}
\label{fig13}
\end{center}
\end{figure}

The similar conclusions can be also drawn from the comparison of
country distributions in the plane $(K,K^*)$ (Fig.~\ref{fig4})
and in the plane $(\tilde{K},\tilde{K}^*)$ (Fig.~\ref{fig13}),
which show data on the same scales. Clearly, the distributions are 
rather different and only for {\it all commodities}
we can see visible correlations (we note that appearance of
ordered short line segments in panels (c,g) around 
$K \approx \tilde{K} \approx  200$ is due to degeneracy of 
$P$ and $\tilde{P}$ values, for those countries which
e.g. do not use {\it barley}, and thus their ordering 
becomes somewhat arbitrary).

Let us discuss in more detail few concrete examples
shown in Tables 1-5 in Appendix. For {\it all commodities}
first 5 positions are very close in both ways of ranking. 
As a significant change we note {\it Canada} which moves from 
$\tilde{K}^*=11$ down to $K^*=16$ and 
{\it Mexico} with respective change from $\tilde{K}^*=13$
to $K^* > 20$: the export of these two countries is
too strongly oriented on {\it USA} that
becomes directly visible through CheiRank
analysis. In contrast {\it Singapore}
moves up from $\tilde{K}^*=15$ to $K^*=11$
that shows the stability and broadness of its 
export trade, a similar situation
appears for {\it India} moving up
from $\tilde{K}^*=19$ to $K^*=12$.

Even more strong changes of ranking appear for
specific commodities. For example for {\it crude petroleum}
Russia moves up from $\tilde{K}^*=2$ to $K^*=1$
showing that its trade network in this product is 
better and broader than the one of {\it Saudi Arabia}.
{\it Iran} moves in opposite direction
from $\tilde{K}^*=5$ down to  $K^*=14$
showing that its trade network is 
restricted to a small number of nearby countries.
A significant improvement of ranking 
takes place for {\it Kazakhstan} moving up
from $\tilde{K}^*=12$ to $K^*=2$.
The direct analysis shows that this happens
due to an unusual fact that {\it Kazakhstan}
is practically the only country
which sells {\it crude petroleum}
to the CheiRank leader in this product {\it Russia}.
This puts {\it Kazakhstan} on the second position.
It is clear that such direction of trade
is more of political or geographical origin
and is not based on economic reasons.

For {\it natural gas} there are also significant
differences between two ways of ranking.
Thus, {\it USA} moves strongly up
from $\tilde{K}^*=10$ to $K^*=1$
due its broad trade network in this product.
{\it Canada} moves down from
$\tilde{K}^*=2$ to $K^*=8$ due to its too strong
trade orientation on {\it USA}. A small country
{\it Trinidad and Tobago} moves up from
$\tilde{K}^*=15$ to $K^*=2$ since it provides 
about 70\% of import of top leader {\it USA}. 

Significant reordering appears also for {\it barley} trade.
Thus, the leader {\it Ukraine} moves down from 
$\tilde{K}^*=1$ to  $K^*=6$ due to too narrow trade network
and {\it USA} moves up from $\tilde{K}^*=8$
to $K^*=3$ due to its broad trade network.

For trade of {\it cars} we have 
{\it France} going up from $\tilde{K}^*=7$ 
to $K^*=3$ due to its broad export network. 
Also {\it Thailand} goes strongly up from
$\tilde{K}^*=19$ to $K^*=10$ due to its broad trade links.
We note that on the side of import we have 
strong change for {\it Nigeria} which
moves from $\tilde{K} > 20$ up to $K=1$.
This is the most populated country in Africa
with a strong income due to oil trade which provides
a large fraction of {\it USA} import. With such  oil
income {\it Nigeria} buys {\it cars} from all over the world
and thus becomes at the top of PageRank.

Finally we note that among top 20 countries 
ranked in 2DRank $K_2$ by {\it all commodities} in 2008
(see Table 1)
there 14 among {\it G-20} major economies \cite{g20}.
At the same time ExportRank gives 13, and 
ImportRank gives 14 countries from 19 of {\it G-20}-list.
We attribute a difference in 5 countries
to political and geographical aspects of {\it G-20}-selection. 

\section{Discussion}

In this work we constructed the Google matrix of the WTN
using the enormous UN COMTRADE database. From this
matrix we obtained PageRank and CheiRank of all
world countries in various types of trade products
for years 1962 - 2009.
This new approach gives a democratic type of ranking
being independent of the trade amount of a given country.
In this way rich and poor countries are treated on equal democratic
grounds.   In a certain sense PageRank probability
for a given country  is proportional to its
rescaled import flows while CheiRank is proportional
to its rescaled export flows inside of the WTN.

The global characteristics of the world trade 
are analyzed on the basis of this new type of ranking.
Even if all countries are treated now on equal democratic grounds 
still we find at the top rank the group of 
industrially developed countries
approximately corresponding to {\it G-20} (74\%).
Our studies establish the existence of two solid state
domains of rich and poor countries which remain
stable during the years of consideration. 
Other countries correspond to a gas phase with 
ranking strongly fluctuating in time. We propose a simple
random matrix model which well describes the statistical properties
of rank distribution for the WTN.

The comparison between usual ImportRank--Export\-Rank 
(see e.g. \cite{cia2010})
and our PageRank--CheiRank approach shows that the later
highlights the trade flows in a new useful manner which
is complementary to the usual analysis. 
The important difference between these two
approaches is due to the fact that  ImportRank--ExportRank
method takes into account only global amount of money
exchange between a country and the rest of the world while
Page\-Rank--CheiRank approach takes into account all links and money
flows between all countries.
We hope that this new approach based on the Google matrix will
find further useful applications to investigation 
of various flows in trade and economy.

{\bf Acknowledgments:}
We thank Gilles Saint-Paul (TSE, Toulouse) for useful discussions
on economics and Olga Chepelianskaia 
(UN Bangkok) for insights on UN databases.
Our special thanks are addressed to 
Arlene Adriano and Matthias Reister
(United Nations Statistics Division) for provided
help and friendly access to the UN COMTRADE database \cite{comtrade}.

\begin{table*}[h]%
\section{Appendix}
\caption {Top 20 ranking for \emph{all commodities} -- 2008.}
\label {table1}\centering %
\begin{tabular}{r*{5}{c}} 
\hline
Ran& $K$&  $K^*$&$K_2$ &$\tilde{K}$ & $\tilde{K}^*$\\
\hline	 
1&         USA  &           China  &          USA  & 	         USA  &           China \\
2&          Germany  &          USA  &           China  & 	          Germany  &           Germany \\
3&          China  &           Germany  &           Germany  & 	          China  &          USA \\
4&          France  &          Japan  &          Japan  & 	          France  &          Japan \\
5&         Japan  &           France  &           France  & 	         Japan  &           France \\
6&         UK  &          Italy  &          Italy  & 	         UK  &          Netherlands \\
7&         Italy  &          Russian Fed.  &          UK  & 	         Netherlands  &          Italy \\
8&         Netherlands  &          Rep. of Korea  &          Netherlands  & 	         Italy  &          Russian Fed. \\
9&         India  &          UK  &          India  & 	          Belgium  &          UK \\
10&         Spain  &          Netherlands  &          Rep. of Korea  & 	          Canada  &           Belgium \\
11&          Belgium  &          Singapore  &           Belgium  & 	         Spain  &           Canada \\
12&          Canada  &          India  &          Russian Fed.  & 	         Rep. of Korea  &          Rep. of Korea \\
13&         Rep. of Korea  &           Belgium  &           Canada  & 	         Russian Fed.  &          Mexico \\
14&         Russian Fed.  &           Australia  &          Spain  & 	         Mexico  &          Saudi Arabia \\
15&         Nigeria  &           Brazil  &          Singapore  & 	         Singapore  &          Singapore \\
16&         Thailand  &           Canada  &          Thailand  & 	         India  &          Spain \\
17&         Mexico  &          Spain  &           Australia  & 	         Poland  &          Malaysia \\
18&         Singapore  &          South Africa  &           Brazil  & 	         Switzerland  &           Brazil \\
19&         Switzerland  &          Thailand  &          Mexico  & 	         Turkey  &          India \\
20&          Australia  &          U. Arab Emir.  &          U. Arab Emir.  & 	          Brazil  &          Switzerland \\
\hline
\end{tabular}
\end{table*}

\begin{table*}[h]%
\caption {Top 20 ranking for \emph{crude petroleum} -- 2008.}
\label {table2}\centering %
\begin{tabular}{r*{5}{c}} 
\hline
Ran& $K$&  $K^*$&$K_2$ &$\tilde{K}$ & $\tilde{K}^*$\\
\hline
1&         USA  &          Russian Fed.  &          USA  & 	         USA  &          Saudi Arabia \\
2&          Canada  &          Kazakhstan  &          India  & 	         Japan  &          Russian Fed. \\
3&         Netherlands  &          U. Arab Emir.  &          Singapore  & 	          China  &          U. Arab Emir. \\
4&          Belgium  &          USA  &          UK  & 	         Italy  &          Nigeria \\
5&         India  &           Ecuador  &          South Africa  & 	         Rep. of Korea  &          Iran \\
6&          China  &          Saudi Arabia  &           Canada  & 	         India  &          Venezuela \\
7&          Germany  &          India  &           Australia  & 	          Germany  &          Norway \\
8&         Japan  &          South Africa  &          U. Arab Emir.  & 	         Netherlands  &           Canada \\
9&         Rep. of Korea  &          Nigeria  &           Colombia  & 	          France  &            Angola \\
10&         UK  &          Sudan  &           Azerbaijan  & 	         UK  &          Iraq \\
11&         Singapore  &           Azerbaijan  &          Malaysia  & 	         Spain  &          Libya \\
12&         Italy  &          Venezuela  &           Brazil  & 	         Singapore  &          Kazakhstan \\
13&          Australia  &          Norway  &           Belgium  & 	          Canada  &          Kuwait \\
14&         Malaysia  &          Iran  &          Trinidad and Tobago  & 	         Thailand  &           Azerbaijan \\
15&         Spain  &            Algeria  &           France  & 	          Belgium  &            Algeria \\
16&          France  &          Singapore  &          Netherlands  & 	          Brazil  &          Mexico \\
17&          Brazil  &          Kuwait  &          Kenya  & 	         Turkey  &          UK \\
18&         Sweden  &          UK  &            Angola  & 	         South Africa  &          Qatar \\
19&         South Africa  &            Angola  &           China  & 	         Poland  &          Oman \\
20&         Thailand  &           Canada  &          Thailand  & 	          Australia  &          Netherlands \\
\hline
\end{tabular}
\end{table*}

\begin{table*}[h]%
\caption {Top 20 ranking for \emph{natural gas} -- 2008.}
\label {table3}\centering %
\begin{tabular}{r*{5}{c}} 
\hline
Ran& $K$&  $K^*$&$K_2$ &$\tilde{K}$ & $\tilde{K}^*$\\
\hline
1&         USA  &          USA  &          USA  & 	         Japan  &          Norway \\
2&         Japan  &          Trinidad and Tobago  &           France  & 	         USA  &           Canada \\
3&         Rep. of Korea  &          Norway  &           Belgium  & 	          France  &            Algeria \\
4&         Spain  &          UK  &          South Africa  & 	         Rep. of Korea  &          Russian Fed. \\
5&          France  &          Russian Fed.  &          Italy  & 	         Spain  &          Qatar \\
6&         Italy  &          Oman  &           Canada  & 	          Belgium  &           Belgium \\
7&         Nigeria  &           Australia  &          UK  & 	         UK  &          Indonesia \\
8&          China  &           Canada  &          Malaysia  & 	         Italy  &          Malaysia \\
9&         Poland  &           France  &           Germany  & 	          Germany  &          Netherlands \\
10&         Portugal  &            Algeria  &           China  & 	         Ukraine  &          USA \\
11&          El Salvador  &          South Africa  &          Nigeria  & 	         Netherlands  &           Australia \\
12&         Kenya  &          Kazakhstan  &           Greece  & 	         Mexico  &          Nigeria \\
13&          Belgium  &          Qatar  &          Turkey  & 	          China  &          Saudi Arabia \\
14&         Guatemala  &          Saudi Arabia  &          Kenya  & 	         India  &          U. Arab Emir. \\
15&          Germany  &          U. Arab Emir.  &          Netherlands  & 	         Hungary  &          Trinidad and Tobago \\
16&         Mexico  &           Belgium  &          Rep. of Korea  & 	          Czech Rep.  &           Germany \\
17&          Ecuador  &          Pakistan  &          Spain  & 	          Canada  &          Oman \\
18&         Malaysia  &          Singapore  &          Russian Fed.  & 	          Brazil  &           Egypt \\
19&         South Africa  &          Netherlands  &          India  & 	         Turkey  &          UK \\
20&         Slovenia  &          Italy  &          Japan  & 	         Thailand  &          Turkmenistan \\
\hline
\end{tabular}
\end{table*}

\begin{table*}[h]%
\caption {Top 20 ranking for \emph{barley} -- 2008.}
\label {table4}\centering %
\begin{tabular}{r*{5}{c}} 
\hline
Ran& $K$&  $K^*$&$K_2$ &$\tilde{K}$ & $\tilde{K}^*$\\
\hline
1&         Saudi Arabia  &           France  &          USA  & 	         Saudi Arabia  &          Ukraine \\
2&         Yemen  &           Canada  &           Germany  & 	          Germany  &           France \\
3&          Germany  &          USA  &          Netherlands  & 	         Japan  &           Australia \\
4&         U. Arab Emir.  &           Australia  &           Denmark  & 	          China  &           Canada \\
5&         USA  &           Germany  &          Italy  & 	          Belgium  &           Germany \\
6&         Israel  &          Ukraine  &           Belgium  & 	         Netherlands  &          Russian Fed. \\
7&         Japan  &          Rep. of Moldova  &           China  & 	         Syria  &           Argentina \\
8&         Netherlands  &          UK  &          U. Arab Emir.  & 	         Iran  &          USA \\
9&         Oman  &           Argentina  &          UK  & 	         Jordan  &           Denmark \\
10&          Greece  &          Spain  &          Spain  & 	         USA  &          Kazakhstan \\
11&         Italy  &           Denmark  &          Singapore  & 	          Denmark  &          Romania \\
12&          Croatia  &          Kazakhstan  &          Rep. of Korea  & 	         Italy  &          UK \\
13&         Syria  &          Netherlands  &          Malaysia  & 	         Tunisia  &          Hungary \\
14&         Kuwait  &          Russian Fed.  &          Russian Fed.  & 	         Israel  &          Spain \\
15&          Cyprus  &          India  &           Austria  & 	          Colombia  &           Bulgaria \\
16&          Denmark  &          Hungary  &          Poland  & 	           Algeria  &          Netherlands \\
17&         Occ. Palestinian Terr.  &          Romania  &           Brazil  & 	         Kuwait  &          Lithuania \\
18&         Switzerland  &           Belgium  &          Ireland  & 	          Brazil  &          Sweden \\
19&          Bosnia Herzegovina  &          Lithuania  &           France  & 	         Morocco  &           Belgium \\
20&         Jordan  &          Sweden  &          South Africa  & 	         Turkey  &          India \\
\hline
\end{tabular}
\end{table*}

\begin{table*}[h]%
\caption {Top 20 ranking for \emph{cars} -- 2008.}
\label {table5}\centering %
\begin{tabular}{r*{5}{c}} 
\hline
Ran& $K$&  $K^*$&$K_2$ &$\tilde{K}$ & $\tilde{K}^*$\\
\hline
1&         Nigeria  &           Germany  &           Germany  & 	         USA  &           Germany \\
2&          Germany  &          Japan  &          USA  & 	          Germany  &          Japan \\
3&          France  &          USA  &           France  & 	         UK  &          USA \\
4&         USA  &          Rep. of Korea  &          UK  & 	          France  &           Canada \\
5&         Russian Fed.  &           France  &           Belgium  & 	         Italy  &          Rep. of Korea \\
6&         UK  &          UK  &          Spain  & 	         Russian Fed.  &          UK \\
7&          Belgium  &           Belgium  &          Italy  & 	          Belgium  &           France \\
8&         Ukraine  &          Spain  &          Japan  & 	          Canada  &          Spain \\
9&         Italy  &          South Africa  &           Australia  & 	         Spain  &           Belgium \\
10&          Greece  &          Thailand  &           Canada  & 	          China  &          Mexico \\
11&         Venezuela  &          Mexico  &           China  & 	         Netherlands  &          Italy \\
12&         Spain  &          Italy  &          Netherlands  & 	          Australia  &          Slovakia \\
13&          China  &           Canada  &          U. Arab Emir.  & 	         U. Arab Emir.  &           Czech Rep. \\
14&         Netherlands  &          U. Arab Emir.  &           Austria  & 	         Saudi Arabia  &          Poland \\
15&          Australia  &           Czech Rep.  &          South Africa  & 	          Austria  &          Sweden \\
16&         Japan  &          Slovakia  &          Poland  & 	         Mexico  &          Turkey \\
17&           Albania  &          Hungary  &          Thailand  & 	         Poland  &          Hungary \\
18&         Romania  &           Australia  &          Russian Fed.  & 	         Switzerland  &           Austria \\
19&         Sudan  &           Austria  &          Turkey  & 	          Finland  &          Thailand \\
20&          Canada  &           China  &          Portugal  & 	         Ukraine  &          South Africa \\
\hline
\end{tabular}
\end{table*}

\end{document}